\documentclass[aps,prl,groupedaddress,floats,showpacs,floatfix, reprint]{revtex4-1}

\usepackage{amsmath}
\usepackage{amssymb}

\usepackage{tikz}
\usetikzlibrary{matrix,shapes,arrows,positioning,chains}

\usepackage{xcolor}
\usepackage{graphicx}
\usepackage{caption}
\usepackage{subfigure}
\usepackage{bm}
\usepackage{hyperref}

\begin{document}

\title{Superconductivity in the Uniform Electron Gas: Irrelevance of Kohn-Luttinger Mechanism}

\author{Xiansheng Cai}
\author{Tao Wang}
\affiliation{Department of Physics, University of Massachusetts, Amherst, MA 01003, USA}

\author{Nikolay V. Prokof'ev}
\email{prokofev@physics.umass.edu}
\affiliation{Department of Physics, University of Massachusetts, Amherst, MA 01003, USA}
\author{Boris V. Svistunov}
\email{svistunov@physics.umass.edu}
\affiliation{Department of Physics, University of Massachusetts, Amherst, MA 01003, USA}
\affiliation{Wilczek Quantum Center, School of Physics and Astronomy, Shanghai Jiao Tong University, Shanghai 200240, China}

\author{Kun Chen}
\email{kunchen@flatironinstitute.org}
\affiliation{Center for Computational Quantum Physics, Flatiron Institute, 162 5th Avenue, New York, New York 10010}
	
\date{\today}
\begin{abstract}
 We study the Cooper instability in jellium model in the controlled regime of 
 small to intermediate values of the Coulomb parameter $r_s \leq 2$.  
 We confirm that superconductivity naturally emerges from purely repulsive interactions 
 described by the Kukkonen-Overhauser vertex function. By employing 
 the implicit renormalization approach {and the discrete Lehmann representation} we reveal that even in the small-$r_s$ limit, the dominant mechanism behind Cooper instability is based on 
 dynamic screening of the Coulomb interaction---accurately captured by the random phase approximation, whereas the Kohn-Luttinger
 contribution is negligibly small and, thus, not relevant. 
  
\end{abstract}
		
\maketitle



\textit{Introduction.}\textemdash
Conventional BCS theory predicts Cooper instability at low temperature in a Fermi liquid 
with weak short range attractive coupling between the low-energy electrons, typically
originating from the electron-phonon interaction (EPI). 
With the discovery of unconventional superconducting systems, such
as $d$-wave cuprate superconductors \cite{bednorzPossibleHighTcSuperconductivity1986,wuSuperconductivity93New1987,chuSuperconductivity150HgBa2Ca2Cu3O81993,tsueiPairingSymmetryCuprate2000}, 
$s^{+-}$ $\textrm{Fe}$-based superconductors \cite{IronbasedLaFeAs,mazinUnconventionalSuperconductivitySign2008,wangFunctionalRenormalizationGroupStudy2009,siStrongCorrelationsMagnetic2008}, 
multi-layer graphene systems \cite{uchoaSuperconductingStatesPure2007,nandkishoreChiralSuperconductivityRepulsive2012,caoUnconventionalSuperconductivityMagicangle2018,zhouSuperconductivityRhombohedralTrilayer2021},
etc., alternative theories of Cooper instability
have been drawing great research interest. In the last decades, a number 
of different mechanisms were proposed for the paring instability in systems where 
the EPI alone was not sufficient to explain the data \cite{kohnNewMechanismSuperconductivity1965,takadaPlasmonMechanismSuperconductivity1978,rietschelRoleElectronCoulomb1983,ruhmanSuperconductivityVeryLow2016,monthouxSuperconductivityPhonons2007}.
Some of them are based on purely repulsive bare electron-electron interactions
\cite{kohnNewMechanismSuperconductivity1965,takadaPlasmonMechanismSuperconductivity1978,rietschelRoleElectronCoulomb1983}. 
In this work, we revisit superconducting properties of the uniform electron gas
(jellium model) where pairing instability originates from purely repulsive 
Coulomb interpaticle interactions and Fermi energy is the only 
relevant energy scale (all other energy scales are emergent). Our prime focus 
is the quantitative study of two canonical scenarios emerging from renormalized 
interactions: the Kohn-Luttinger (KL) mechanism based on the $2k_F$ singularity (where $k_F$ is the Fermi momentum) and the dynamic screening mechanism.

In 1965, Kohn and Luttinger argued that for any weak short-range repulsive interaction, the two-particle effective interaction induced by many-body effects always becomes attractive at large enough orbital momenta $\ell \gg 1$, and could lead to Cooper instability \cite{kohnNewMechanismSuperconductivity1965}. They used the same analysis to estimate an effective Cooper channel coupling for the static screened Coulomb interaction.
{The KL mechanism has motivated a series of theoretical attempts to explain unconventional superconducting systems} \cite{baranovSuperconductivitySuperfluidityFermi1992,chubukovKohnLuttingerEffectInstability1993,galitskiKohnLuttingerPseudopairingTwodimensional2003,gonzalezKLTBG}, and is widely believed to be the dominant mechanism leading to superconductivity in jellium at small $r_s$. 
However, a thorough investigation of the dominant mechanism leading to superconductivity in this system is still missing: 
(i) The KL approach is based on the static screened potential, which   
is an uncontrolled approximation; (ii) We are not aware of any precise  
numerical study of the KL mechanism because it only emerges at large $\ell$ and leads to extremely small values of $T_c$; (iii) Whenever Cooper instability is observed in the simulation, one needs to differentiate between different scenarios behind it by revealing and evaluating their contributions separately. Therefore, whether the KL mechanism ever becomes dominant in the uniform electron gas in the high-density limit $r_s\rightarrow0$ is still an unsolved fundamental question.

It has been known for decades that dynamic screening of the Coulomb interaction could also induce the paring instability in jellium.
Early work by Tolmachev \cite{Tolmachev1958} demonstrated that even 
if the Cooper channel coupling is repulsive at all 
frequencies, after its high-frequency part is renormalized to a 
smaller value the net result might be an attractive low-frequency effective potential.
Later, Takada and others calculated the critical temperatures $T_c$ of jellium numerically using various forms of dynamically screened Coulomb interaction 
\cite{takadaPlasmonMechanismSuperconductivity1978,rietschelRoleElectronCoulomb1983,takadaSuperconductivityOriginatingRepulsive1989,takadaPwavePairingsDilute1993}.
Subsequent studies also reported that dynamic screening 
plays important role in the superconductivity of metallic hydrogen and alkali metals \cite{richardsonEffectiveElectronelectronInteractions1997b,richardsonHighTemperatureSuperconductivity1997b} as well as in the dilute electron gas \cite{ruhmanSuperconductivityVeryLow2016}. The superconducting phase diagram produced by Takada \cite{takadaPwavePairingsDilute1993} stated that jellium enters a normal phase at $r_s<2.0$, in contradiction with the KL prediction. Apparently, the values of $T_c$ at large $\ell$ were too small to be resolved and, therefore, were ignored for practical purposes. Because of this limitation, the KL and dynamic screening mechanisms 
have never been quantitatively compared to each other, despite their 
coexistence in the uniform electron gas. 

To determine which mechanism is dominant and under what conditions, 
we study the Cooper instability in the controlled regime of small 
to intermediate values of $r_s \leq 2$. The
particle-particle irreducible four-point vertex is approximated with the
Kukkonen-Overhauser (KO) ansatz \cite{kukkonenElectronelectronInteractionSimple1979a}, which becomes exact in the high-density limit $r_s \rightarrow 0$. We compare contributions from both mechanisms in two different ways. In the first protocol, 
we compute the largest eigenvalues, $\lambda(T)$, of the gap equation down to $T/E_F = 10^{-6}$ (below we use Fermi energy $E_F$ as the unit of energy). 
We then remove the $q=2k_F$ singularity in the polarization function, on which the KL mechanism is based, and measure the relative change of the eigenvalue, $\eta(T)=\delta \lambda /\lambda$. Finally, we estimate the magnitude of $\eta(T_c)$, which represents the relative contribution of the KL mechanism. This perturbative treatment is justified if $\eta(T)$ is small, which turns out to be 
always the case.
The second protocol is based on the implicit renormalization (IR) approach \cite{chubukovImplicitRenormalizationApproach2019a}.  
By integrating out the high-frequency/energy degrees of freedom
the IR approach solves a new eigenvalue problem for which the largest eigenvalue, $\bar{\lambda}(T)$, is also equal to unity at $T=T_c$. 
The crucial advantage of looking at $\bar{\lambda}(T)$ instead of $\lambda (T)$ is that its temperature dependence is a linear function of $\ln (T)$ 
for a properly chosen energy separation scale, and, thus, can be accurately extrapolated to $T_c$ from $T \gg T_c$. Computational costs are further dramatically reduced by employing the discrete Lehmann representation (DLR) \cite{kayeDiscreteLehmannRepresentation2021a}.
This combination of methods is what allows us to determine the superconducting channel $\ell_c$ with the highest value of $T_c$ at $r_s=0.33,0.5,1,2$; otherwise, the problem cannot be solved using standard techniques.   
By computing the critical values of orbital channels, $\ell_{KL} (r_s)$, 
when the KL mechanism first induces an attractive Cooper channel coupling,
and comparing the asymptotic behavior of $\ell_c$ and $\ell_{KL}$ at small $r_s$ we determine what mechanism is dominating in the high-density limit. 

Within the first protocol we find the KL mechanism contribution remains extremely 
small, and therefore irrelevant, for any value of $r_s$ when the transition 
temperatures exceed $10^{-10^{6}}$. 
The second protocol reveals that the Cooper instability in jellium takes place at all values of $r_s$ tested. More importantly, the dominant channel $\ell_c$ increases much slower than the critical channel of the KL mechanism, $\ell_{KL}$, as $r_s$ decreases, indicating that in the high-density limit superconductivity is induced by the dynamic screening effects accurately captured by the random phase approximation, 
well before the KL mechanism could have any impact.  

\textit{Model.}\textemdash
Jellium model is defined by the Hamiltonian
\begin{equation}
  \label{eq:hamiltonian}
  H = \sum\limits_{\mathbf{k} \sigma} \epsilon_{\mathbf{k}} a_{\mathbf{k} \sigma}^\dagger a_{\mathbf{k} \sigma}^{\:}
  + {\textstyle{1 \over 2}}\!\!\!\!\!\!\!\sum^{\mathbf{q}\neq 0}_{\mathbf{q}, \mathbf{k}, \mathbf{k^\prime}, \sigma , \sigma^\prime}\!\!\!\!\!
  V_{q}\,  a_{\mathbf{k}+\mathbf{q} \sigma}^\dagger a_{\mathbf{k^\prime}-\mathbf{q} \sigma^\prime}^\dagger a_{\mathbf{k^\prime}^{\:} 
  \sigma^\prime}a_{\mathbf{k} \sigma}^{\:},
\end{equation}
with $a_{\mathbf{k} \sigma}^\dagger$ the creation operator of an 
electron with momentum $\mathbf{k}$ and spin $\sigma=\uparrow,\downarrow$,
dispersion $\epsilon_{k}=\frac{k^2}{2m_e}-\mu$, and Coulomb potentail 
$V_{q}= \frac{4\pi e^2}{ q^2}$. 
The dimensionless coupling parameter 
(the Wigner-Seitz radius) is given by $r_s=\frac{1}{a_0} (\frac{3}{4\pi n})^{\frac{1}{3}}$, where n is the number density, and $a_0$ is the Bohr radius. 
The gap function equation reads:
\begin{equation}
  \label{eq:eigenDelta}
  \lambda(T) \Delta_{\omega_n,\mathbf{k}}
  =
  -T\sum\limits_{m}\int\frac{d\mathbf{p}}{{(2\pi)^d}}
  \Gamma_{\omega_m,\mathbf{p}}^{\omega_n,\mathbf{k}} G_{\omega_m,\mathbf{p}}G_{-\omega_m,-\mathbf{p}}\Delta_{\omega_m,\mathbf{p}}.
\end{equation}
Here $\Gamma$ is the particle-particle irreducible four-point vertex, $G$ is the
single particle Green's function, $\Delta$ is the gap function, and $\lambda$ is its eigenvalue. The key approximation used in this work is the 
Kukkonen-Overhauser ansatz \cite{kukkonenElectronelectronInteractionSimple1979a,supmat} 
for $\Gamma$:
\begin{align}
V_{\vec{\sigma}\vec{\sigma}'}^{KO}(\omega, q)= V_q&+V_+(q)^2 Q_+(\omega,q)\nonumber\\
&+V_-(q)^2 Q_-(\omega,q) \vec{\sigma} \cdot \vec{\sigma}' ,
\end{align}
with
\begin{align}
Q_\pm(\omega,q)=-\frac{\Pi_0(\omega,q)}{1+V_\pm(\omega,q)\Pi_0(\omega,q)},\\
V_+=(1-G_+)V,\quad V_-=-G_-V .
\end{align}
It is defined in terms of the polarization function
$\Pi_0$ (based on the convolution of bare Green's functions) and local field factors $G_\pm(q)$. 
When $G_\pm(q)$ are set to zero, the KO interaction reduces to the random phase approximation (RPA). The local field factors encode the many-body exchange and correlation effects beyond RPA. For direct comparison with previous work by Takada we adopt the same ansatz for $G_\pm(q)$   \cite{takadaSuperconductivityOriginatingRepulsive1989,takadaPwavePairingsDilute1993}, and take the functional form of $\Pi_0$ to be that at $T=0$:
\begin{align}
		\Pi_{0}(q, \omega) \simeq \frac{mk_F}{2\pi^2} &P\left( \frac{q}{2k_F}, \frac{m\omega}{qk_F} \right),\\
		P(z,u) = 1+\frac{1-z^2+u^2}{4z}&\ln\frac{{(1+z)}^2+u^2}{{(1-z)}^2+u^2} \nonumber\\&- u \tan^{-1}\frac{2u}{u^2+z^2-z}.
\end{align}
This is justified by the smallness of the critical temperature.
The gap equation is decomposed into different orbital channels, 
$\ell$, which are solved independently. 
The critical temperature $T_c$ in each channel corresponds to the point where the largest eigenvalue $\lambda(T)$ equals unity. 
{For every choice of the vertex function
considered in this work the single particle self-energy was computed within the $G_0W_0$ approximation as in    \cite{takadaPwavePairingsDilute1993}.} 

\begin{figure}[htbp]
	\subfigure[]{\label{Fig:KL_diagrams:1}
	\includegraphics[width=.35\linewidth]{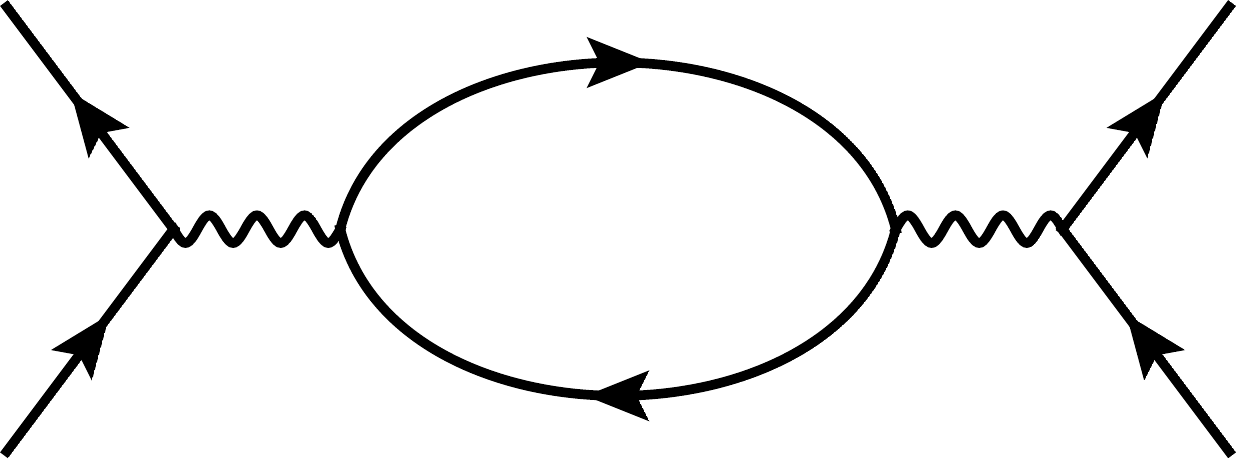}}
	\subfigure[]{\label{Fig:KL_diagrams:2}
	\includegraphics[width=.29\linewidth]{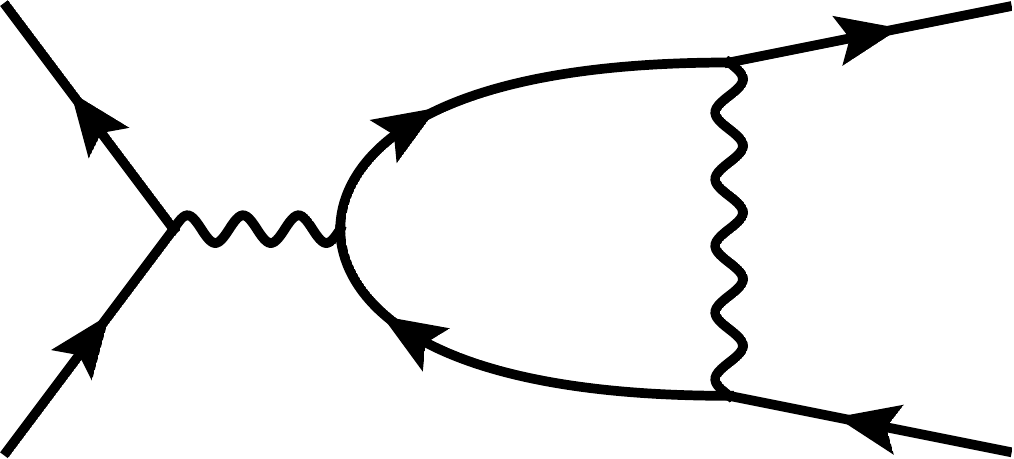}}
	\subfigure[]{\label{Fig:KL_diagrams:3}
	\includegraphics[width=.29\linewidth]{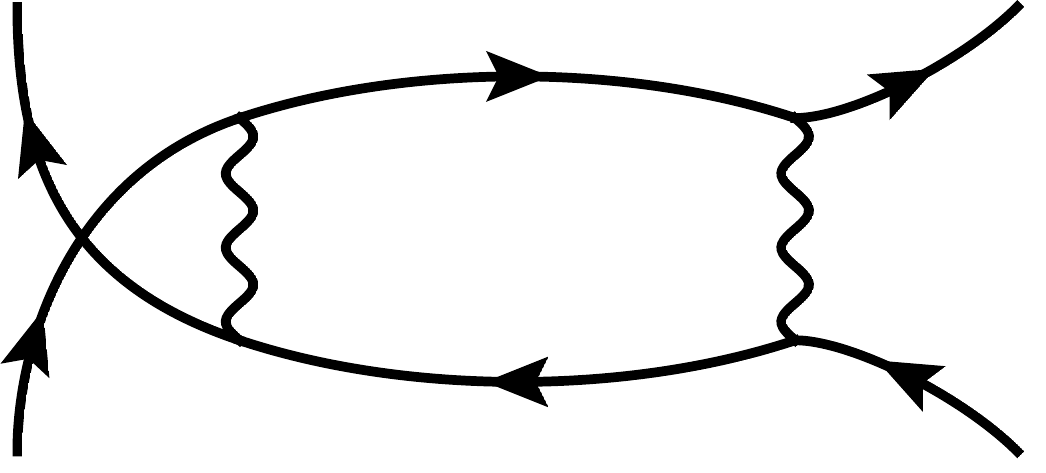}}
	\caption{\label{Fig:KL_diagrams} Second-order diagrams contributing to the particle-particle irreducible four-point vertex $\Gamma$ in the KL analysis. Only the first bubble diagram is resummed within the  RPA approach, while all three diagrams are resummed in the KO formulation. In the small-$r_s$ limit, the contributions of the diagrams (b) and (c) might prove appreciable only if the Kohn-Luttinger mechanism is the leading channel of Cooper instability; 
	otherwise, the leading contribution is accurately captured by RPA.}
\end{figure}


\textit{Kohn-Luttinger mechanism.}\textemdash
Cooper instability in the KL theory is induced by the logarithmic singularity in the static effective interaction at $q=2k_F$  \cite{kohnNewMechanismSuperconductivity1965}. Within the 
second-order perturbation theory it arises from diagrams shown in Fig.~\ref{Fig:KL_diagrams}. 
After projecting static $\Gamma$ to the $\ell$-th orbital 
channel,
\begin{equation}
W_{\ell}(k, p, \omega=0)=\!\! \int_{-1}^{1}  P_{\ell}(\chi)\,  \Gamma\left(k,p,\chi, \omega=0 \right) d \chi, 
\end{equation}
where $\chi = \cos \theta$,  with $\theta$ the angle between momenta $\mathbf{k}$ and $\mathbf{p}$, and $P_{\ell}(\chi)$ are Legendre polynomials,
Kohn and Luttinger found that in the large-$\ell$ limit, $W_{\ell}(k_F, k_F, \omega=0)$ decays as $\ell^{-4}$ and oscillates between the odd and even values of $\ell$.
The attractive effective coupling at large enough odd $\ell$ could then give rise to Cooper instability. 

The KL treatment silently ignores the dynamic nature of screening 
despite the fact that at any finite frequency the Coulomb potential cannot be screened at small momenta $q \ll \omega /v_F$. 
Moreover, singular nature of the Coulomb potential at small $q$ calls for proper resummation of diagrams shown in Fig.~\ref{Fig:KL_diagrams} beyond the second-order perturbation theory.
Clearly, a more controlled analysis is necessary to quantitatively evaluate the relative importance of KL and dynamic screening effects in the jellium model. 

\begin{figure}[htbp]
	\centering
	\includegraphics[width=.9\linewidth]{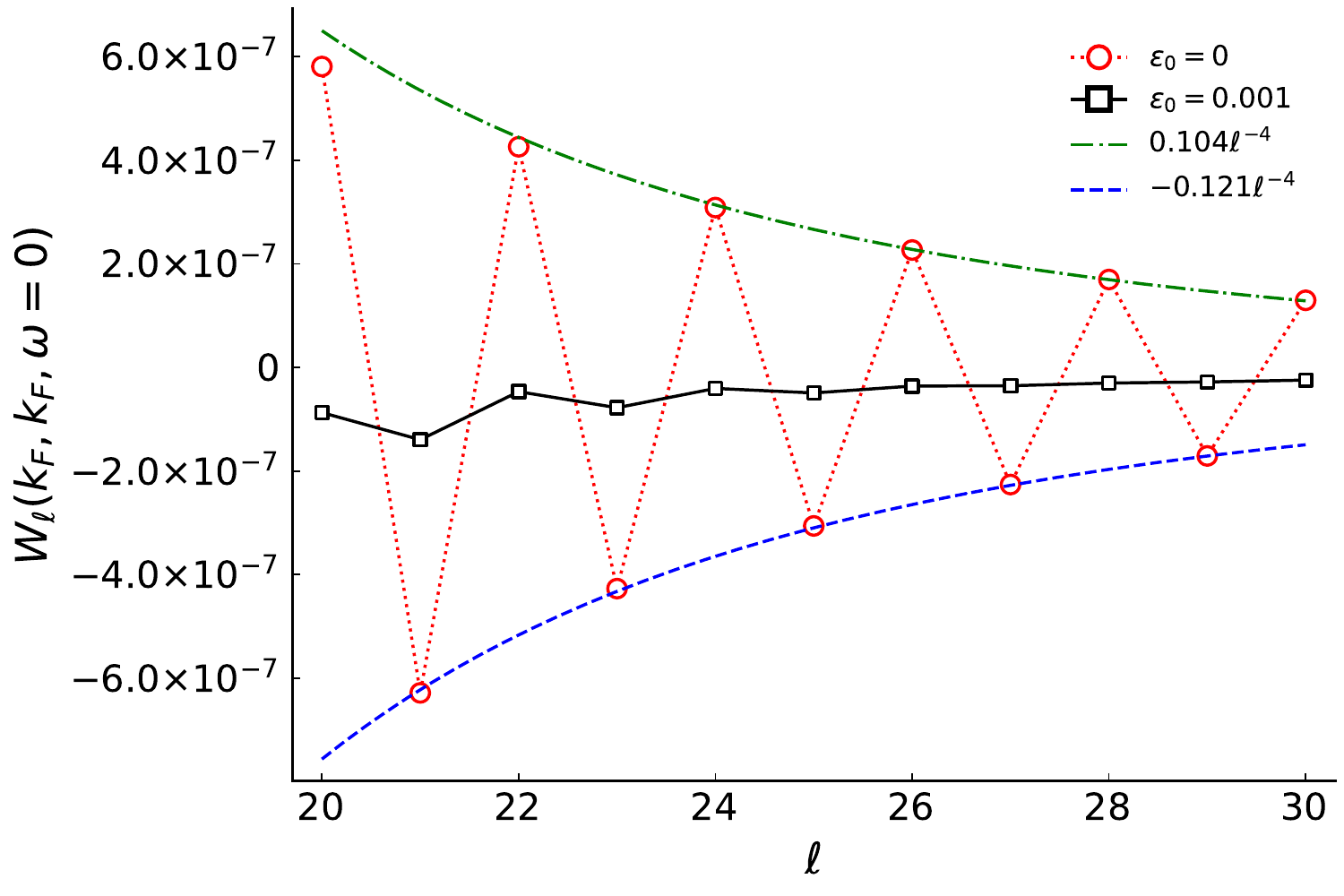}
	\caption{\label{fig:Keff} The static KO vertex function on the Fermi surface $W_{\ell}(k_F, k_F, \omega=0)$ at $r_s = 1.0$.
		Dotted red line with circles represents the original $W_{\ell}$ with clear KL oscillations decaying 
		as $\ell^{-4}$ (dot-dash green and dashed blue fits). 
		The solid black line with squares represents the regularized $W_{\ell}$.}
\end{figure}

Resummation of diagrams shown in Fig.~\ref{Fig:KL_diagrams} is 
achieved within the KO vertex function that provides an excellent framework
for thorough investigation of competing mechanisms. To quantify the contribution of the $q=2k_F$ singularity, we introduce regularized $\Pi_\epsilon (q,\omega)$ that differs from $\Pi_\epsilon (q,\omega)$ by the replacement ${(1-z)}^2+u^2 \to {(1-z)}^2+u^2+\epsilon (z)$ with $\epsilon (z) = \epsilon_0 e^{-4(z-1)^2}$. 
%
%
This modification is limited to the vicinity of 
the $2k_F$ singularity and ensures that changes in $\lambda$ 
are attributed to the KL mechanism. 
The value of $\epsilon_0=0.001$ was chosen by establishing when 
$W_{\ell}(k_F, k_F, \omega=0)$ for the regularized KO interaction  
has its odd-even channel oscillations suppressed, see Fig.~\ref{fig:Keff}. 
Next, we study the effect of $\epsilon_0$ on the gap equation 
eigenvalues $\lambda_{\ell}$.  

\begin{figure}[htb]
  \centering
  \includegraphics[width=.9\linewidth]{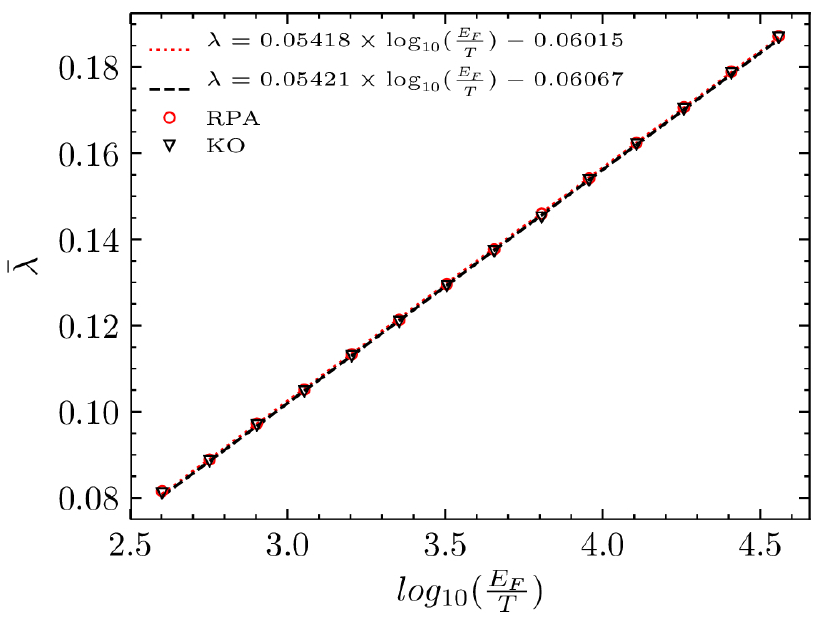}
  \caption{\label{fig:flow} Temperature dependence of the 
  eigenvalues $\bar{\lambda}(T)$ for RPA (red circles) and KO (black triangles) vertex functions at $r_s=2.0$ and $\ell=3$. 
  The linear fits of the RPA (red dotted line) and KO (black dashed line) data are almost identical. 
  {The extrapolated value of $T_c = 2.71\times10^{-20}E_F$ is extremely small.}}
\end{figure}

\textit{Implicit renormalization approach.}\textemdash
It is impossible to solve the gap equation directly when the critical temperature is extremely low because the number of required Matsubara frequency points is too large. Thus, reliable extrapolation from  
$T\gg T_c$ is essential for determining where $\lambda(T)=1$.
Such an extrapolation can hardly be done for frequency-dependent vertexes
because $\lambda(T)$ turns out to be an unknown nonlinear function of $\ln T$. 
The implicit renormalization (IR) approach proposed in 
Ref.~\cite{chubukovImplicitRenormalizationApproach2019a} 
offers a solution to this problem. The idea is to decompose 
the gap function into two complementary parts, $\Delta=\Delta^{(1)}+\Delta^{(2)}$, with $\Delta_n^{(1)}=0$ for $|\omega_n|>\Omega_c$, and $\Delta_n^{(2)}=0$ for $|\omega_n|<\Omega_c$, and solve an eigenvalue problem for the low-energy part $\Delta_n^{(1)}$ only. [The integration 
of high-energy degrees of freedom with the IR protocol 
is achieving the same goal as the pseudopotential theory].
The new eigenvalue $\bar{\lambda}(T)$ is expected to have a 
nearly perfect linear dependence on $\ln T$ for a properly chosen energy scale separation. The IR approach allows
us to accurately determine $T_c$ as low as ${10}^{-20}E_F$ by extrapolating $\bar{\lambda}(T)$ from the $10^{-5}< T/E_F<10^{-3}$ interval. Once the Tolmachev-McMillan logarithm \cite{Tolmachev1958,morelCalculationSuperconductingState1962,rietschelRoleElectronCoulomb1983} is accounted for, the linear flow
of $\bar{\lambda}(T)$ illustrated in Fig.~\ref{fig:flow} provides direct access to the Coulomb pseudopotential $\mu^*$ in a given orbital channel $\ell$. 
Note that the difference between $\mu^*$ of the KO and RPA vertex functions is smaller than a few percent for odd $\ell$ at $r_s\leq 2.0$ and even $\ell$ for $r_s\leq 1.0$. This indicates that higher-order vertex corrections for considered values of $r_s$ are negligible.

\textit{Discrete Lehmann representation.}\textemdash
 Even within the IR protocol, solving the gap equation at exponentially low temperature faces technical challenges because the vertex function is a multi-dimensional 
 object with nontrivial structure in momentum and frequency. 
 The key step is optimization of frequency grids to store just enough information for accurate interpolation of functions. Fortunately, for a given ultraviolet cutoff 
 $\omega_{\rm max}$ and numerical accuracy $\epsilon$, the required 
 grids are provided by the recently developed discrete Lehmann representation \cite{kayeDiscreteLehmannRepresentation2021a, kayeDiscreteLehmannRepresentation2022a}, or DLR. 
 The number of grid points scales as $O(\ln(\omega_{\rm max}/T)\ln(\frac{1}{\epsilon}) )$, and 
 only 65 frequencies are required to achieve accuracy $\epsilon=10^{-10}$ at $\omega_{\rm max}/T=10^{5}$. 
 Fast implementation of all key operations including Fourier transforms, interpolation, and convolution are available within the DLR\cite{supmat}.
 By dramatically reducing memory and computational costs,  
 the DLR grids allow us to simulate much lower temperatures and compute
 extremely small $T_c$. Specifically, we were able to determine 
 the dominant superconducting orbital channel $\ell_c$ at $r_s\geq 0.33$.

\begin{figure}[htbp]
  \centering
 {
  \includegraphics[width=.9\linewidth]{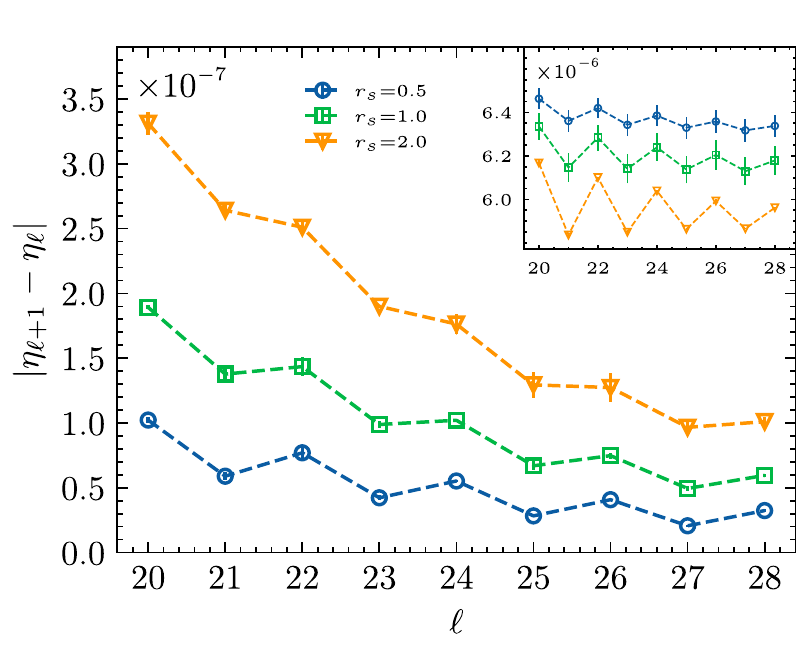}
	}
  \caption{\label{fig:kl_contribution_ko} 
 Amplitudes of the relative eigenvalue oscillations, $|\eta_{\ell+1}-\eta_{\ell}|$,
 induced by regularization of the KO vertex at $T=10^{-5}E_F$ and $r_s=$ 0.5 (blue circles), $r_s=1$ (green squares), and $r_s=2$ (yellow triangles). Inset:  relative eigenvalue changes $\eta_{\ell}$. Data in the main figure and insert 
 involve scaling factors of $10^{-7}$ and $10^{-6}$, respectively.}
\end{figure}

\textit{Results.}\textemdash In the  
Fig.~\ref{fig:kl_contribution_ko} inset we show the relative change of eigenvalues $\eta=\delta \lambda / \lambda$ at $T=10^{-5}E_F$ when the KO function is regularized. We observe that KL oscillations are superimposed on a slowly decaying background
originating from the vertex modification in finite vicinity of the $q=2k_F$ point.
The contribution from singularity is best characterized by the oscillation amplitude, 
or the difference between the $\eta_\ell$ and $\eta_{\ell+1}$ values plotted in the 
lower panel of Fig.~\ref{fig:kl_contribution_ko}. 
We observe that (i) the relative contribution of the KL mechanism is $\sim 10^{-7}$; 
(ii) the oscillation is less pronounced for smaller $r_s$.
Both facts indicate that the KL mechanism is irrelevant for superconducting properties of jellium. 

\begin{figure}[htbp]
	\centering
	\includegraphics[width=.9\linewidth]{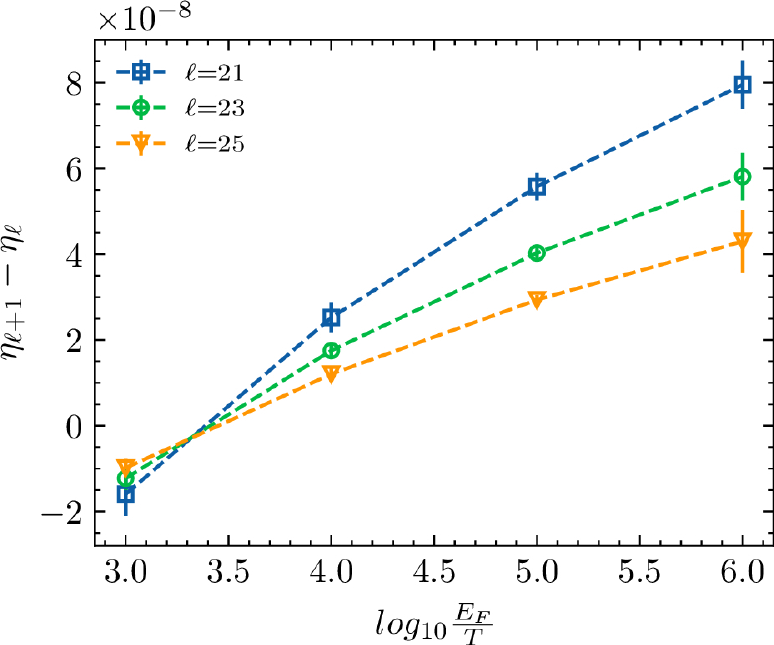}
	\caption{\label{fig:kl_Temp} Oscillation amplitudes as functions of $\log_{10}(E_F/T)$ for $r_s=0.5$ and $\ell=21$ (blue squares), $\ell=23$ (green circles) and $\ell=25$ (yellow triangles). When $T/E_F$ is reduced by one order of magnitude, the amplitudes increase by less then $3 \times10^{-8}$ for $\ell \geq 21$ at $T \sim 10^{-6} E_F$. }
\end{figure}

Our conclusions do not change with temperature as evidenced by
simulation results shown in Fig.~\ref{fig:kl_Temp}. 
The KL contributions $\eta_{\ell+1}-\eta_\ell$ increase when temperature decreases, 
but the rate is tiny and the curves tend to saturate. 
Even if $\eta_{\ell+1}-\eta_\ell$ were to grow at a constant rate
beyond {$T/E_F=10^{-6}$}, they would be smaller than a few percent at {$T/E_F=10^{-10^{6}}$}, which is ``zero" for all practical purposes.  

\begin{figure}[htbp]
	\centering
	\includegraphics[width=.9\linewidth]{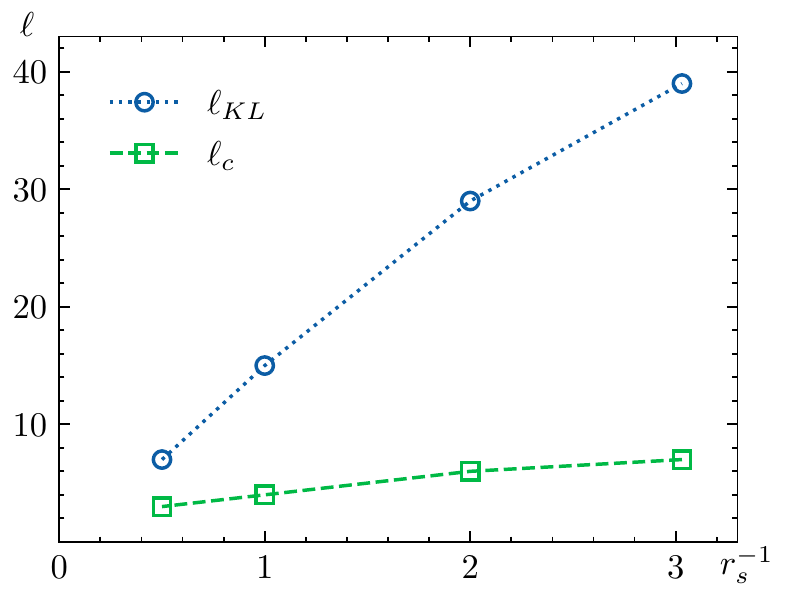}
	\caption{\label{fig:l_KL} Dominant superconducting channel $\ell_c$
	(green squares connected by the dashed line) as a function of $r_s$ along with the critical channel of the KL mechanism $\ell_{KL}$ (blue circles connected by the dotted line). }
\end{figure}

Academically speaking, i.e. regardless of how small $T_c$ is in the $r_s \to 0$ limit, 
the KL mechanism is not ruled out by results presented above because 
we cannot determine whether $\eta_{\ell+1}-\eta_\ell$ ultimately saturate to small values that decrease with $r_s$. However, recall that the KL mechanism is attractive only at $\ell \geq \ell_{KL}$, where $\ell_{KL}(r_s)$ is the first channel with $W_{\ell}(k_F, k_F, \omega=0) < 0$. It has to be compared with the IR solution for the dominant superconducting channel, $\ell_c (r_s)$, for dynamic KO vertex function 
(this can be done for $1/3 \leq r_s\leq 2$). The comparison presented in Fig.~\ref{fig:l_KL} demonstrates that $\ell_{KL}>\ell_c$ for all $r_s$ and the difference  keeps growing when $r_s \to 0$. Thus, superconductivity in the 
$\ell_c$ channel is induced by the dynamical screening well before the 
KL mechanism becomes viable, including the  $r_s\rightarrow 0$ limit.  

\textit{Conclusions.}\textemdash 
We studied superconductivity in the jellium model
for $r_s\leq2$ by solving the gap equation based on the Kukkonen-Overhauser vertex function. We find that superconductivity emerges from repulsive Coulomb interactions
due to dynamic screening effects. The Kohn-Luttinger mechanism, often 
assumed to be the prime reason behind Cooper instability in the high-density limit, 
is not relevant for two reasons: (i) For $r_s\leq2$, it contributes to the Cooper instability only at orbital momenta $\ell \geq \ell_{KL}$ much larger than the dominant superconducting channel $\ell_c$ selected by the dynamic screening effect; 
(ii) For $\ell \geq \ell_{KL}$, the
relative contribution of the KL mechanism is extremely small numerically, and can be safely ignored. Since at $r_s\leq 1$ the physics of Cooper instability is accurately captured by RPA, one may attempt to find an analytic solution in the $r_s \to 0$ limit. 

We solve the fundamental problem of Cooper instability in jellium 
in the high-density limit, and revise a popular, yet incorrect, 
belief that the $q=2k_F$ singularity is the key reason. Our approach 
offers a systematic way for studies of the Cooper instability in other 
correlated electronic systems.

\begin{acknowledgements}
The authors thank C.~Kukkonen for discussions on connections between effective interactions and superconductivity, and Kristjan Haule for discussions 
and support of exchange visits. KC thanks Y.~Deng, B.~Wang and P.~Hou for helpful discussions. NP, TW, and XC acknowledge support by the the National Science Foundation under the grant DMR-2032077. BS and KC acknowledge support by the Simons Collaboration on the Many Electron Problem. The Flatiron Institute is a division of the Simons Foundation. {The symmetrized discrete Lehmann representation algorithm is a registered Julia package: \href{https://github.com/numericalEFT/Lehmann.jl}{https://github.com/numericalEFT/Lehmann.jl}}\cite{kayeDiscreteLehmannRepresentation2022a}.
\end{acknowledgements}

%

\end{document}



\title{Supplemental Materials: Superconductivity in the Uniform Electron Gas: Irrelevance of Kohn-Luttinger Mechanism}
\author{Xiansheng Cai}
\author{Tao Wang}
\affiliation{Department of Physics, University of Massachusetts, Amherst, MA 01003, USA}

\author{Nikolay V. Prokof'ev}
\email{prokofev@physics.umass.edu}
\affiliation{Department of Physics, University of Massachusetts, Amherst, MA 01003, USA}
\author{Boris V. Svistunov}
\email{svistunov@physics.umass.edu}
\affiliation{Department of Physics, University of Massachusetts, Amherst, MA 01003, USA}
\affiliation{Wilczek Quantum Center, School of Physics and Astronomy, Shanghai Jiao Tong University, Shanghai 200240, China}

\author{Kun Chen}
\email{kunchen@flatironinstitute.org}
\affiliation{Center for Computational Quantum Physics, Flatiron Institute, 162 5th Avenue, New York, New York 10010}

\maketitle

\section{Kukkonen-Overhauser theory for the homogeneous Coulomb gas}

To properly resum the singular behavior at momentum transfer $q=2k_F$ 
that ultimately leads to the Kohn-Luttinger mechanism, one needs 
to consider geometric series for the bubble and exchange diagrams 
on equal footing. 
\begin{figure}[h]
  \centering
  \includegraphics[width=0.9\columnwidth]{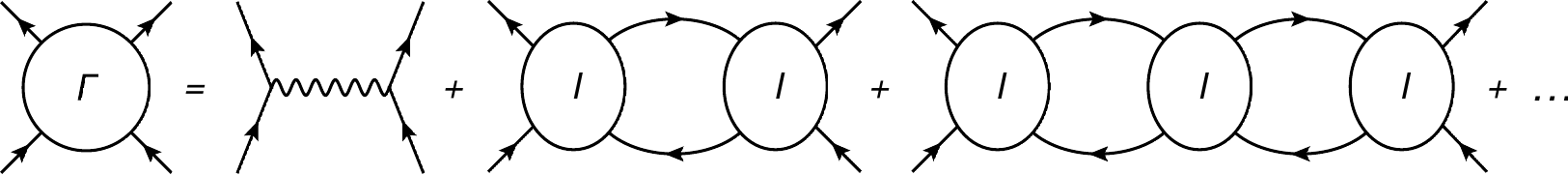}      
  \caption{\label{fig:ko_diag}
  Resummation of the particle-hole channel.
  }
\end{figure}
Now, apart from the bare potential, the series are built on the vertex
$I$ which contains all diagrams irreducible in the particle-hole channel.

Since the relevant part of all particle-hole irreducible diagrams has
the same sigularity, the assumption is that the relevant diagrams have 
the same functional form as in RPA.
Thus, instead of computing the full particle-hole irreducible vertex 
$I$ an approximation is used where $I$ contains the bare potential and the
$G_{\pm}$ terms originating from exchange diagrams, while its functional form is that of an effective potential:
\begin{eqnarray}
\label{eq:ph_ko}
  I \approx V-V(G_+ + G_- \vec{\sigma} _1 \cdot \vec{\sigma} _2).
\end{eqnarray}
This scheme accounts for all relevant singular contributions 
as long as the $G_{\pm}$ functions are chosen appropriately. 

Equation (\ref{eq:ph_ko}) leads to the standard expression for 
the KO vertex, and provides an opportunity to take full advantage 
of the well-studied ansatz for local field factors.
In our simulations we employ the same ansatz as in Ref.~\cite{takadaPwavePairingsDilute1993,takadaSuperconductivityOriginatingRepulsive1989}.
It is somewhat outdated for large values of $r_s$, but is more than adequate for $r_s \leq 2$ considered in our work when 
the difference between the KO and RPA treatments is very small. 

\section{ Discrete Lehmann representation for the gap-function equation}

Since our goal is compact, accurate, and efficient processing 
of the frequency space, we omit momentum variables here for brevity.
The eigenvalue problem to solve is given by Eq.(2) in the main text
with the four-point vertex $\Gamma$ approximated by $V_{KO}(\omega)=V+W(\omega)$, where $V$ is the Coulomb potential and $W(\omega)$ is the frequency dependent part to be represented by the 
DLR. The gap-function $\Delta$ and the anomalous Green's function 
$F=G G \Delta$ are also assumed to be compatible with DLR for asymmetric correlators:
\begin{eqnarray}
W_m &=& \sum_{\alpha} W_{\alpha} K_{c}(m, \alpha), \\
F_n &=& \sum_{\alpha} F_{\alpha} K_{a}(n, \alpha), \\
\Delta_n &=& \sum_{\alpha} \Delta_{\alpha} K_{a}(n, \alpha),
\end{eqnarray}
with
\begin{eqnarray}
K_{c}(m, \alpha) &=& \frac{2\alpha}{\alpha^2+\omega_m^2}(1-e^{-\alpha \beta}), \\
K_{a}(n, \alpha) &=& \frac{2\alpha}{\alpha^2+\omega_n^2}(1+e^{-\alpha \beta}).
\end{eqnarray}
Substituting these expressions in Eq.(2) in the main text we get 
\begin{eqnarray}
\nonumber
&\lambda&\Delta_n = -T\sum_{m}[V+W_{n-m}] F_m \\*
\nonumber
&=&-TV\sum_{\alpha}F_{\alpha}\sum_{m}K_a(m, \alpha) \\*
   &&-T\sum_{\alpha,\gamma}W_{\gamma}F_{\alpha}\sum_{m}K_a(m, \alpha)K_c(n-m, \gamma).
\end{eqnarray}
With pre-computed coefficients $S_{\alpha}=\sum_{m}K_a(m, \alpha)$ and 
$C_{n,\gamma,\alpha}=\sum_{m}K_a(m, \alpha)K_c(n-m, \gamma)$, 
the right hand side of the gap equation involves compact sums over 
$\alpha$ and $\gamma $. 
Moreover, simple analytical structure of $K_c$ and $K_a$
allows one to compute $S_\alpha$ and $C_{n, \gamma,\alpha}$  
analytically:
\begin{eqnarray}
    \nonumber
    S_{\alpha} &=& \sum_{m}K_{a}(m,\alpha) \\*
    \nonumber
    &=& \sum_{m} \frac{2\alpha}{\alpha^2+\omega_n^2}(1+e^{-\alpha \beta})\\*
    &=& (1+e^{-\alpha \beta}) \tanh(\alpha\beta/2)\label{eq:asw}
\end{eqnarray}
and
\begin{eqnarray}
    \nonumber
    &&C_{n,\gamma,\alpha} = \sum_{m}K_a(m, \alpha)K_c(n-m, \gamma) \\*
    \nonumber
    &=& \sum_{m}\frac{4\alpha\gamma(1+e^{-\alpha \beta})(1-e^{-\gamma \beta})}
    {(\alpha^2+\omega_m^2)(\gamma^2+(\omega_n-\omega_m)^2)} \\*
    \nonumber
    &=& \frac{4\alpha\gamma(1+e^{-\alpha \beta})(1-e^{-\gamma \beta})}
    {{(\alpha^2-\gamma^2)}^2+\omega_n^4+4\omega_n^2{(\alpha^2+\gamma^2)}}\\*
    \nonumber
    &&[\frac{\alpha^2-\gamma^2+\omega_n^2}{2\gamma}\frac{1+e^{-\alpha\beta}}{1-e^{-\alpha\beta}}\\*
    &&+\frac{\gamma^2-\alpha^2+\omega_n^2}{2\alpha}\frac{1-e^{-\gamma\beta}}{1+e^{-\gamma\beta}}]\label{eq:csw}
\end{eqnarray}

\section{Implicit renormalization scheme and its compatibility with the discrete Lehmann representation}

In the IR approach the gap-function is split into the low- and high-energy parts using projection operators 
$P_m = \Theta(\Omega-|\omega_m|)$ and $\bar{P}=1-P$:
\begin{eqnarray}
\Delta^{(L)}_n &=& P_n\Delta_n, \\
\Delta^{(H)}_n &=& \bar{P}_n \Delta_n.
\end{eqnarray}
The modified eigenvalue problem is formulated as
\begin{eqnarray}
\bar{\lambda} \Delta^{(L)}_n &=& -P_nT\sum_{m}[V+W_{n-m}]F_m, \\
\Delta^{(H)}_n &=& -\bar{P}_nT\sum_{m}[V+W_{n-m}]F_m.
\end{eqnarray}
The problem is that sharp cutoffs in $\Delta^{(L)}$ and $\Delta^{(H)}$ cannot not be represented by the compact DLR. 
However, the right hand sides of these equations is compatible with DLR. Thus instead of $\Delta^{(L)}$ and $\Delta^{(H)}$ we store ''complete versions" $\Delta^{(1)}$ and $\Delta^{(2)}$, and solve
\begin{eqnarray}
\bar{\lambda} \Delta^{(1)}_n &=& -T\sum_{m}[V+W_{n-m}]F_m, \\
\Delta^{(2)}_n &=& -T\sum_{m}[V+W_{n-m}]F_m,
\end{eqnarray}
with $F=P F^{(1)} + (1-P)F^{(2)}$ calculated from the anomalous propagators based on $\Delta^{(1)}$ and $\Delta^{(2)}$.
The main computational cost now looks as: 
\begin{eqnarray}
\nonumber&& -T\sum_{m}[V+W(\omega_n-\omega_m)]F(\omega_m) = 
\qquad \qquad \;\;\;\;\;\;\;\;\;\;\;
\\*
\nonumber&&-TV\sum_{\alpha}F_{1,\alpha}\sum_{m}P(\omega_m)K_a(m, \alpha) \\*
\nonumber&&-TV\sum_{\alpha}F_{2,\alpha}\sum_{m}\bar{P}(\omega_m)K_a(m, \alpha) \\*
\nonumber&&-T\sum_{\alpha,\gamma}W_{\gamma}F_{1,\alpha}\sum_{m}P(\omega_m)K_a(m, \alpha)K_c(n-m, \gamma) \\*
&&-T\sum_{\alpha,\gamma}W_{\gamma}F_{2,\alpha}\sum_{m}\bar{P}(\omega_m)K_a(m, \alpha)K_c(n-m, \gamma). 
\end{eqnarray}
Again, with predetermined coefficients describing sums over Matsubara frequencies we only need to handle compact sums. 
For the low-energy part these sums are done numerically; 
the high-energy part is obtained by subtracting the low-energy part 
from the analytic expressions (\ref{eq:asw}) and (\ref{eq:csw}).